\documentclass[conference]{IEEEtran}
\IEEEoverridecommandlockouts
\usepackage{cite}
\usepackage{amssymb,amsmath,amsfonts,latexsym,mathtext}
\usepackage{algorithmic}
\usepackage{graphicx}
\usepackage{textcomp}
\usepackage{xcolor}
\usepackage[utf8]{inputenc}
\def\BibTeX{{\rm B\kern-.05em{\sc i\kern-.025em b}\kern-.08em
    T\kern-.1667em\lower.7ex\hbox{E}\kern-.125emX}}
\begin{document}

\title{Analysis Of Congestion Control In Data Channels With Frequent Frame Loss\\
\thanks{This report contains results of the research project supported by Russian Foundation for Basic Research, grants no. 18-07-01109, 16-47-330055}
}

\author{\IEEEauthorblockN{Yuri Monakhov}
\IEEEauthorblockA{\textit{Department of Informatics and Information Security} \\
\textit{Vladimir State University}\\
Vladimir, Russia \\
unklefck@gmail.com}
\and
\IEEEauthorblockN{Anna Kuznetsova}
\IEEEauthorblockA{\textit{Department of Informatics and Information Security} \\
\textit{Vladimir State University}\\
Vladimir, Russia \\
akuznecova@vlsu.ru}
}

\maketitle

\begin{abstract}
Development of optimal control procedures for congested networks is a key factor in maintaining efficient network utilization. The absence of congestion control mechanism or its failure can lead to the lack of availability for certain network segments, and in severe cases -- for the entire network. The paper presents an analytical model describing the operation of the TCP Reno congestion control algorithm in terms of differential calculus and queuing systems. The purpose of this research is to explore the possibilities and ways of increasing the virtual channel capacity utilization efficiency in a lossy environment.
\end{abstract}


\section{Introduction}
Due to the active development of information processes, the increase in the scale of the distribution of wireless networks and the widespread development and use of the Internet of Things (IoT) methodology, the volume of critical data transmitted through wireless networks is enormously increasing. The main problem is that the scale of wireless networks is growing quantitatively, by adding new data transfer nodes to existing ones, often without complex positioning of the equipment with respect to the architecture of the premises. This approach leads to a significant qualitative reduction in the data transfer rate, which in turn is reflected in significant congestion of the data transmission channel and routing equipment, high network packet delays and frequent packet losses. This approach increases the percentage of tasks that will not be executed in the required time. It will inevitably affect the profits of organizations. At the same time, the organization does not have enough time and resources to change and redevelop the network architecture without putting business processes on a long halt.

Therefore, there is an actual task to research and develop prototypes for new data transfer algorithms in the TCP/IP stack. Such algorithms should be able to provide an optimal data transfer rate, to adapt to the resulting overloads and losses in wireless networks. Moreover, they should be independent in the sense of being platform- and architecture-agnostic. This will allow organizations to reduce downtime in the process of using wireless telecommunications networks without significant and expensive replanning of the existing network infrastructure.

In this paper authors research the algorithms for congestion control in a virtual data channel with loss of frames. In such channels there is frequent network packet loss and high latency time on the hop between the source and the router. Congestion control algorithms typically allow adjusting the data rate by increasing or decreasing the amount of "unconfirmed" data segments transmitted in one conditional round based on various virtual transmission channel metrics. The study of algorithms for preventing congestions is usually performed as investigating the effectiveness of such algorithms, which one could see in the works of such researchers as Veres and Boda [1], Lakshman [2], Chen [3], Raman [4], Lee [5], Genin [6], Bonald [7] and others [8-11].The purpose of our research is to increase bandwith utilization efficiency in a virtual transmission channel by developing a new adaptive congestion control algorithm in data transmission networks under conditions of long delays and frequent loss of frames. In this paper we present an analytical model that makes it possible to comprehensively simulate the work of the congestion control algorithm based on the theory of differential equations and queuing systems. This model is characterized by the simplicity of the experiments, as well as the ability to determine the optimal parameters of the network system, which directly affect the data transfer process under various initial conditions and virtual data link states. 

\section{Preliminaries}

\subsection{Virtual channel as the research object}
Tanenbaum [12] provides us with the following definition of a virtual data transmission channel: it is a communication channel in a packet routing network that connects two or more subscribers. The virtual channel consists of consecutive physical links in the data transmission system between the communication nodes (switches and routers). It also includes the logical and physical links inside the switches and routers on the path between subscribers.

Thus, the main components of such a channel are: a data source, a data receiver, a routing device, and an unstable transmission medium in which significant frame losses are observed. In this case, data integrity should be preserved, and successful transmission from the source to the receiver should be guaranteed. Therefore the TCP protocol is used as the transport protocol.

The data transfer medium with loss of frames has the following main features [13]:

1) Frequent loss of network data packets.

2) The presence of long random delays in data transmission.

3) Dependence of the maximum bandwidth of the channel, the number of frames lost and the length of delays from the physical location of the receiver and the transmitter relative to the router.

The source and receiver in the virtual channel are represented by the same type of client devices. Their data transmitters are capable of operating in the above-described data communication environment. Information exchange between the source and the receiver is carried out according to the "point-to-point" method. Client devices run an operating system that implements the TCP/IP stack for data transfer. The router is a network device that supports the processing and routing of network packets over IP and has a network packet processing buffer.

The definition of the concept of "congestion" in data transmission networks is formulated in the works of a number of authors [14-17]. The most close to this study is Chiang in [18]: congestion in the TCP/IP flow (TCP flow) is the state of the data link between the two links, in which the total (aggregate) amount of incoming data per unit time (bps - bits per second) exceeds the bandwidth of the data link between these links and/or exceeds the amount of data that the network equipment can handle.
Graphically, the case of network channel congestion can be represented using the following image (Fig. 1). Two TCP streams are sent to the buffer of router R. An aggregated stream of 3 MB/s overloads the router's buffer, and it begins to drop packets.

\begin{figure}[htbp]
\centerline{\includegraphics[width=9cm]{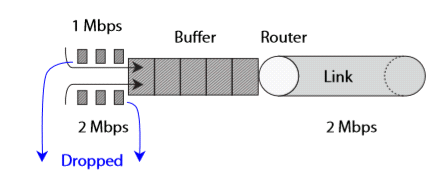}}
\caption{Schematic representation of network channel congestion.}
\label{fig1}
\end{figure}

To avoid such a situation, the TCP/IP stack contains congestion control mechanisms that allow the source of the data stream to regulate the information sending speed with respect to the bandwidth of the data link.
\subsection{TCP Reno congestion control scheme} 
We denote the congestion avoidance process as a strategy of adapting the data rate by a TCP stream source. According to this srategy, the source will decrease or increase the transmitted data quantity per unit time based on a certain control algorithm. As a result, the essence of the congestion control process in a TCP protocol reduces to changing the \textit{cwnd} parameter based on certain rules. In this case, the main events signaling the presence of congestion in the data transmission channel are [19]:
\begin{itemize}
\item packet loss event,
\item duplicate ACK receive event.
\end{itemize}
According to RFC 2581, the main strategies used by the TCP protocol to change the size of the congestion window are as follows:
\begin{itemize}
\item Slow Start
\item Congestion Avoidance
\item Fast Retransmit
\item Fast Recovery
\end{itemize}
All the techniques described above are used by the TCP protocol to control the size of the overload window [20]. Most modern operating systems use the TCP Reno algorithm as the default overload prevention algorithm.
Let's present the mode of TCP Reno operation in the form of a block-module diagram outlining its main steps:

\begin{figure}[htbp]
\centerline{\includegraphics[width=9cm]{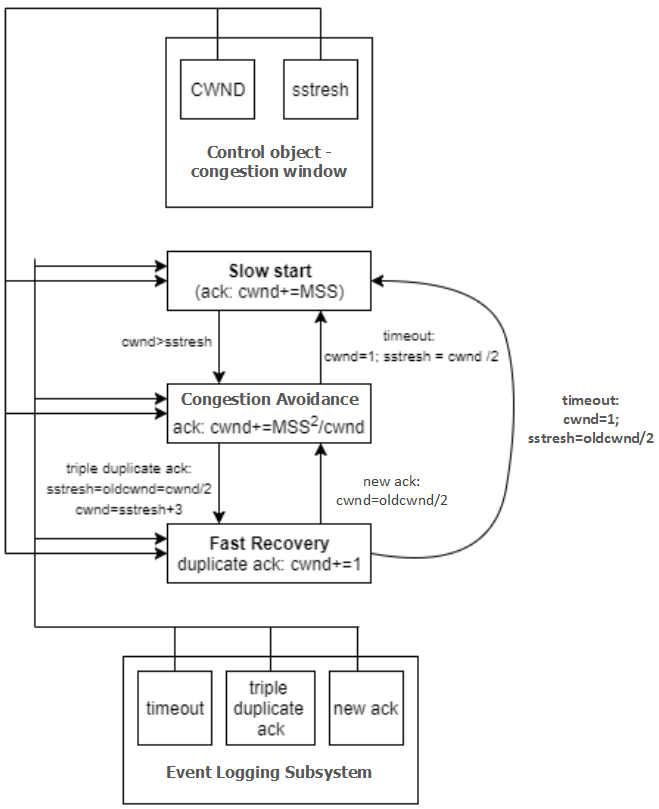}}
\caption{TCP Reno block-module diagram.}
\label{fig2}
\end{figure}

\section{Problem statement}
Let $w(t)$ be the value of the congestion window size at time $t$.
Then the phase of "congestion avoidance" of the TCP Reno algorithm can be described using the following differential equation:
\begin{equation} 
\frac{dw(t)}{dt} = \frac{(1-p(x(t-T)))}{T}-\frac{(x(t-T)p(x(t-T))w(t))}{2},
\end{equation}

where $T$ – round-trip time (RTT);

$x(t)$ – data transfer rate at time $t$, defined on the range of values $E(x) = [0, C]$, where $C$ is the maximum channel capacity (packets per second);

$p(x(t))$is the packet loss probability as a function of the data transfer rate defined on the domain $E(p) = [0,1]$ for any values of $x(t)\in{[0,C]}$.

From RFC 2001 we get the following equation for the change in the data transfer rate:
\begin{eqnarray} 
\frac{dx(t)}{dt} = \frac{(1-p(x(t-T)))}{T^2}-\nonumber\\-\frac{(x(t-T)p(x(t-T))x(t))}{2},
\end{eqnarray}
The limitation of this model is the absence of a function simulating the initial "slow start" phase of the TCP Reno algorithm.
  
Thus, the problem of investigating the effectiveness of the TCP Reno congestion control algorithm can be formulated as follows:

Let the function $x(t) =(w(t))$ describe the data transfer rate regulated by the TCP Reno algorithm. The function $x(t)$ is defined in some $\sigma$-neighborhood of the point $t_n$, where $\sigma>0$. Then the problem of investigating the effectiveness of the TCP Reno algorithm reduces to finding points of the local maximum $t_n$, i.e. those points for which, given all $t\not= t_n$ belonging to the domain $(t_n-\sigma,t_n+\sigma)$ , the following inequalities hold:
\begin{equation}
\begin{cases}
x(t)\le{x(t_n)} \\
P(x(t))\le{p(x(t_n))}
\end{cases}
\end{equation}

\section{Finding the local maximum points of the TCP Reno data rate function}
For a complex study of the function described above, it is necessary to introduce an equation describing the dependence $p(x(t))$ -- the probability of packet loss -- on the data transfer speed.
A "classic" model was used in development of the TCP Reno algorithm, implying that the packet loss event is associated with a buffer overflow of routing equipment.
There are several mathematical models describing the probability of packet loss depending on the applied algorithm of active router queue management (AQM). At the moment, the most common models are Drop Tail and RED (Random early detection).
The principle of the Drop Tail algorithm is to drop incoming packets  when the specified value of $Q_{max}$, indicating the maximum length of the router queue, is exceeded. There are two types of mathematical models describing the given behavior of the AQM algorithm [8]: a router with a small queue size $(Q_{max}<100)$ and a router with an infinitely expandable queue size $ (Q_{max}\to\infty, Q_{init}>100)$.
 
The principle of the RED algorithm is the early dropping of packets arriving at the network equipment before the device queue is full. In this case, the strategy for dropping packets depends on the scheduler algorithm used.

In this paper we decided to use the Drop Tail algorithm, because this AQM strategy has a simple description and allows a more complex study of the actual data rate equation without additional corrections to the description of the function $p(x(t))$, which derive from using a specific scheduler in the RED model.
 
Let us describe a function that determines the probability of packet loss depending on the data rate in the form of a flow model for both kinds of Drop Tail AQM.
For the Drop Tail AQM algorithm with a small queue size $(Q_{max} <100)$, the function $p(x(t))$ takes the following form:
\begin{equation} 
p(x(t)) = \frac{x(t)^B}{C},
\end{equation}
where $x(t)$ is the data rate (in packets per second) at time $t$;
$C$ - channel bandwidth;
$B$ is the maximum size of the routing device queue (in packets).

For the Drop Tail AQM algorithm with an infinitely expandable queue size $(Q_{max}\to\infty,Q_{init}>100)$, the function $p(x(t))$ takes the following form:
\begin{equation} 
p(x(t)) = \frac{1-(\frac{x(t)}{C})(\frac{x(t)}{C})^B}{1-(\frac{x(t)}{C})^{B+1}}
\end{equation}
where $x(t)$ is the data rate (in packets per second) at time $t$;
$C$ - channel bandwidth;
$B$ is the maximum size of the routing device queue (in packets).

To study these patterns, we present a system with the simplest topology consisting of a single source $S$ generating a single TCP data stream $F$ to the receiver $D$ through a routing device $R$ with a maximum buffer size $B$ controlled by the Drop Tail AQM strategy. At the same time, the channel $A\to R$ has the capacity $C$. The channel capacity  $R\to B$ is much larger than $C$, therefore it can be neglected. Thus, we obtain a classical topology with a single bottleneck, through which data is transmitted using a single TCP stream. The diagram of this model is shown in Fig. 3.

\begin{figure}[htbp]
\centerline{\includegraphics[width=9cm]{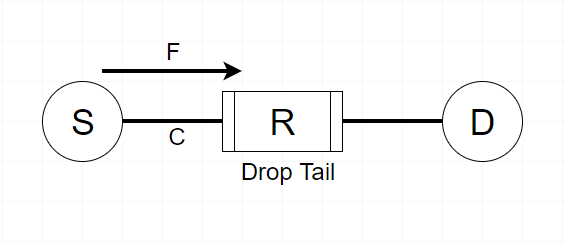}}
\caption{Drop Tail model diagram.}
\label{fig3}
\end{figure}

\begin{figure*}
\centerline{\includegraphics[width =0.9\linewidth]{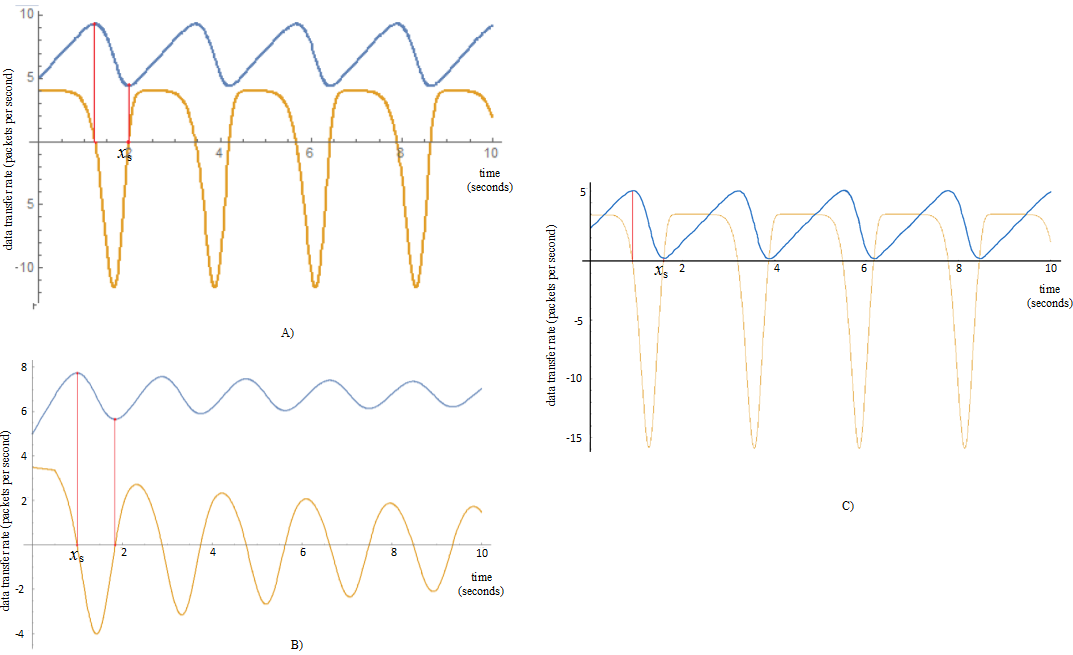}}
\caption{Change in the data transfer rate: A) using data set 1; B) using data set 2; C) using data set 3.}
\label{fig4}
\end{figure*}

The following lemma is, in opinion of the authors, characteristic for both models under study.
\newtheorem{Lem}{Lemma}
\begin{Lem}
let the dynamical system described by the following equation:
\begin{eqnarray} 
p(x(t)) = f(x(t))\nonumber\\
x(t)\in (0,C]
\end{eqnarray}
have such a set of points $x_s(t_x), x_s(t_s )\to(0,C], p(x_s(t_s))\to[0,1], t_s\to[0,\infty]$, for which holds $\frac{d x_s(t_s)}{dt} = 0$. 

\end{Lem}

In other words, for any data transfer process, there is a set of points $x_s(t_s)$ for which the change in the transmission rate is zero.
Then the function $x_s(t_s)$ at such points will take the following value: 
\begin{equation} 
x_s (t_s) = \sqrt{\frac{2(1-p(x_s(t_s)))}{p(x_s(t_s))}}\frac{1}{T}
\end{equation}

To confirm the above lemma, we numerically solve equations (4) and (5) with the help of Wolfram Mathematica 11.

Consider 3 situations of the relation of the router buffer size to the channel capacity ($T$ is the round-trip delay time (in seconds); $C$ - bandwidth of the channel (packets per second); $B$ is the maximum size of the routing device queue (in packets): $B=C; C>B; C<B$.

We shall examine the graphs of the functions $x(t), \dot{x}(t)$ for the existence of the points described earlier. The value of the round-trip delay time is assumed equal to 500 ms. Table 1 contains the corresponding values of the parameters listed above for various experiments. In all the experiments, the graphs are constructed for the value $t \in [0, 10]$ seconds.

\begin{table}[htbp]
\caption{Values of experimental parameters.}
\begin{center}
\begin{tabular}{|p{1cm}|p{2cm}|p{2cm}|p{2cm}|}
\hline
No. & C (pps) & B1 (packets) & T (seconds) \\ \hline
1 & 10 & 10 & 0.5 \\ \hline
1 & 10 & 5 & 0.5  \\ \hline
1 & 5 & 10 & 0.5 \\
\hline
\end{tabular}
\end{center}
\end{table}

Figure 4 shows the graphs  of the solutions obtained with data sets 1, 2, and 3 respectively.

Analyzing the obtained solutions, we note that the graphs actually have points $x_s$ in which $\dot{x}(t) = 0$. These points are the points of the local maximum or minimum. This fact confirms the lemma described above. Let's summarize the obtained data in the final table 2.
\begin{table}[htbp]
\caption{The final data obtained from the test data set no. 1.}
\begin{center}
\begin{tabular}{|p{2cm}|p{1.1cm}|p{2cm}|p{2cm}|}
\hline
Buffer size & Amt. of points & Maximum value & Minimum value \\ \hline
C = B = 10 & 9 & 9 & 5 \\ \hline
C = 10, B = 5 & 10 & 7 & 5  \\ \hline
C = 5, B = 10 & 8 & 5 & 1 \\
\hline
\end{tabular}
\end{center}
\end{table}

A similar study was carried out for model (5), implying the presence of routing equipment with an infinitely expandable buffer size. The obtained data also confirm the lemma. In this case, in all the solutions presented, the graph of $x(t)$ has only one local maximum point and an infinite number of local minimum points. In this case periodic oscillations are not observed in the system (5).

The lemma given in this graph was experimentally confirmed using a test setup. Experimental confirmation will be presented in further work.

\section{Conclusions and future work}
Correlating the obtained data with the propositions, we come to the following conclusions:
\begin{itemize}
\item The dynamical system described by equation (6) has the set of singular points $x_s(t_x), x_s(t_s)\in (0, C], p(x_s (t_s))\in [0,1], t\in [0, \infty] $ for which $\frac {d x_s (t_s)} {dt} = 0$.
\item Based on the numerical solution of the differential equations (4) and (5), the presence of these points in the described systems was shown.
\item The indicated points are points of the local maximum or minimum of the function $x(t)$.
\item The system described by equation (5) has only one local maximum point, and the set of points of a non-strict maximum or minimum.
\end{itemize} 
In the future, it deems possible for the authors to propose a theorem describing the effect of the buffer size of the device on the data transfer rate, as well as to devise a number of techniques that allow increasing the data transfer rate in a medium with frame loss. These results can be used to increase the efficiency of using the capacity of a virtual channel operating in a lossy environment.


\begin{thebibliography}{00}
\bibitem{b1} A. Veres, M. Boda, ``The Chaotic Nature of TCP Congestion Control", IEEE INFOCOM 2000,   pp.1715--1723.
\bibitem{b2}T.V. Lakshman, ``The performance of TCP/IP for neworks with high bandwidth-delay product and random loss", IEEE/ACM Transaction of Networking, vol.5, 1997, pp. 336--350.
\bibitem{b3}X. Chen, S. Wong, ``Stability Analysis of RED Gateway with Multiple TCP Reno connections", Proc. Int. Symp. Circuits and Systems, 2010, pp. 1429--1432.
\bibitem{b4} S. Raman. A. Mohan, ``TCP Reno and queue management: local stability and Hopf bifurcation analysis", 52nd IEEE Conference on Decision and Control, 2013, pp. 3299--3305.
\bibitem{b5}H. Lee. Y. Choi, ``The influence of the Large Bandwidth-Delay product on TCP Reno, New Reno and SACK", Proc. Information Networking Conference, Oita, Japan, Feb. 2001, pp. 327–334.
\bibitem{b6}D. Genin, V. Marbukh, ``Do current Fluid Approximation Models capture TCP Instability'', Journal of Research of NIST, vol 98, 2013.
\bibitem{b7}T. Bonald, ``Comparison of TCP Reno and TCP Vegas via Fluid Approximation'', Institute national de recherche en informatique et en automatique, No 3563. p. 34, 1998.
\bibitem{b8}J. Charzinski, R.Lenhert, P.Tran-Gia, ``Providing Quality of Service in Heterogenous Environment", Proceedings of 18th International Telegraphic Congresse, 2003.   
\bibitem{b9}S. Rastogi, S.Srivastava, ``Comparison Analysis of Different Queuing Mechanisms Droptail, RED and NLRED in Dumb-bell Topology", International Journal of Advanced Research in Computer and Communication Engineering, Vol. 3, Issue 4, 2014
\bibitem{b10}C. Dovrolis, R. Prasad, M. Murray, C. Dobrolis, K. Claffy, ``Bandwidth estimation: metrics, measurement techniques, and tools", IEEE Network, vol. 17, no. 6, pp. 27--35, Apr 2003.
\bibitem{b11}A.A. Altyeb, ``Effect of TCP congestion control mechanism on Self-Simillary of Network Traffic",  Automatic Control and Computer Sciences, Vol. 41, No. 2, 2007, pp. 114–117
\bibitem{b12} A. S. Tanenbaum, D. J. Wetherall, ``Computer Networks", University of Washington, p. 361, 2011.
\bibitem{b13}R. Srikant, Lei Ying, ``Communication Networks: An Optimization, Control, and Stochastic Networks Perspective", Cambridge University Press, p. 352, 2014.
\bibitem{b14} V. Jacobson, M.J. Karels, ``Congestion Avoidance and Control", University of Callifornia at Berkeley, p. 21, 1988.
\bibitem{b15}J.R. Ramakrishnan, ``Congestion avoidance in computer networks with a connectionless network layer", SIGCOMM, 1988. 
\bibitem{b16}A. Kuznetsov, ``Congestion Control in Linux TCP", Proceedings of the FREENIX Track: 2002 USENIX Annual Technical Conference, pp. 49--62 
\bibitem{b17}A.Afanasyev, N.Tilley, P.Reiher, L.Kleinrock, ``Host-to-Host Congestion Control for TCP", IEEE Communications Surveys and Tutorials, Vol. 12, No. 3, 2010, pp. 304-342. 
\bibitem{b18}M. Chiang, ``Networked Life: 20 Questions and Answers", Princeton University, p. 503, 2012.
\bibitem{b19}RFC 5681:  https://tools.ietf.org/html/rfc5681
\bibitem{b20}RFC 2001: https://tools.ietf.org/html/rfc2001


\end{thebibliography}
\end{document}